\documentclass{article}

\usepackage{newlfont}
\usepackage{amssymb}
\usepackage{latexsym}
\usepackage[raiselinks=false,colorlinks=true,citecolor=blue,urlcolor=blue,linkcolor=blue,bookmarksopen=true,dvips]{hyperref}

\makeatletter

\def\normalsize{\@setfontsize\normalsize\@xiipt{14.5}%
\abovedisplayskip 12\p@ \@plus 2\p@ \@minus 2\p@%
\belowdisplayskip \abovedisplayskip
\abovedisplayshortskip \z@ \@plus 2\p@%
\belowdisplayshortskip 3.5\p@ \@plus 2\p@ \@minus 2\p@
\let\@listi\@listI}
\def\small{\@setfontsize\small\@ixpt{9}
\abovedisplayskip 11\p@ plus3\p@ minus6\p@
\belowdisplayskip \abovedisplayskip
\abovedisplayshortskip  \z@ plus3\p@
\belowdisplayshortskip  6.5\p@ plus3.5\p@ minus3\p@
\def\@listi{\leftmargin\leftmargini
 \parsep 4.5\p@ plus2\p@ minus\p@ \itemsep 5\p@
            \topsep 9\p@ plus3\p@ minus5\p@}}
\def\footnotesize{\@setsize\footnotesize{11\p@}\xpt\@xpt}
\def\scriptsize{\@setfontsize\scriptsize\@viiipt{9.5}}
\def\tiny{\@setfontsize\tiny\@vipt{7}}
\def\large{\@setfontsize\large\@xivpt{18}}
\def\Large{\@setfontsize\Large\@xviipt{22}}

\normalsize
\newdimen\@bls
\@bls=\baselineskip
\topmargin \z@

\headheight  \z@
\headsep     \z@
\advance\textheight\topskip
\def\section{\@startsection{section}{1}{\z@}%
        {-3ex plus -1ex minus -.2ex}%
        {2ex plus .2ex}{\large\bfseries}}

\newdimen\labelwidthi
\def\normal@labelsep{0.5em}
\labelsep\normal@labelsep
\leftmargini\labelwidthi    \advance\leftmargini\labelsep
\def\setleftmargin#1#2{\settowidth{\@tempdima}{#2}\labelsep\normal@labelsep
  \csname labelwidth#1\endcsname\@tempdima
  \@tempdimb\@tempdima \advance\@tempdimb\labelsep
  \csname leftmargin#1\endcsname\@tempdimb}
\def\@listI{\leftmargin\leftmargini
  \labelwidth\labelwidthi \labelsep\normal@labelsep
  \topsep .35pc
\partopsep\z@ \parsep\z@ \itemsep .3pc
  \listparindent 1em}
\let\@listi\@listI
\@listi

\newskip\eqnlineskip

\newtheorem{thm}{Theorem}[section]
\newtheorem{cor}[thm]{Corollary}

\newtheorem{prop}[thm]{Proposition}

\renewenvironment{thebibliography}[1]
     {\vspace{4pt}\section*{\refname}%
     \frenchspacing\small%
      \@mkboth{\MakeUppercase\refname}{\MakeUppercase\refname}%
      \list{\@biblabel{\@arabic\c@enumiv}}%
           {\settowidth\labelwidth{\@biblabel{#1}}%
            \leftmargin\labelwidth
            \advance\leftmargin\labelsep
     \itemsep 0.28\@bls \@plus 0.1\@bls \@minus 0.1\@bls
            \@openbib@code
            \usecounter{enumiv}%
            \let\p@enumiv\@empty
            \renewcommand\theenumiv{\@arabic\c@enumiv}}%
      \sloppy
      \clubpenalty4000
      \@clubpenalty \clubpenalty
      \widowpenalty4000%
      \sfcode`\.\@m}
     {\def\@noitemerr
       {\@latex@warning{Empty `thebibliography' environment}}%
      \endlist}

  \def\footnoterule{\kern-3\p@
  \hrule \@width 3pc
  \kern 2.6\p@}

\def\ps@plain{\let\@mkboth\@gobbletwo
 \def\@oddhead{\hfill{\sc Kock}\hfill}%
 \let\@evenhead\@oddhead
 \def\@oddfoot{\hfil {\rmfamily\thepage} \hfil}%
 \let\@evenfoot\@oddfoot}

\let\ps@headings\ps@plain

\ps@headings

\font\eightrm=cmr8 scaled\magstep1

\font\eightit=cmsl10

\def\ps@pageone{\let\@mkboth\@gobbletwo
  \def\@oddhead{\vbox{\baselineskip=10pt \parindent 0pt \obeylines
    \eightrm\hfill MFPS 2012\hfill\break}}
  \let\@evenhead\@oddhead
  \def\@oddfoot{\small\slshape
    \def\@tempa{0}
      {\vbox{\baselineskip=10pt \parindent 0pt \obeylines
    \vspace{.1in}\centerline{\eightit 
      Proceedings of the} \centerline{\eightit
      28th Conference on the Mathematical Foundations of Programming Semantics}
      \centerline{\eightit
      Bath, June 2012.}
    }}}%
  \let\@evenfoot\@oddfoot
}

\makeatother

\usepackage{xspace}

\usepackage{texdraw}
\everytexdraw{\drawdim pt \linewd 0.5 \textref h:C v:C \setunitscale 1
  \arrowheadtype t:F \arrowheadsize l:5 w:3
}

\everytexdraw{%
	\drawdim pt \linewd 0.5 \textref h:C v:C
	\setunitscale 1
}

\newcommand{\Onedot}{
  \bsegment
    \move (0 0) \fcir f:0 r:2.5
  \esegment
}
\newcommand{\smalldot}{
  \bsegment
    \move (0 0) \fcir f:0 r:1.5
  \esegment
}

\newcommand{\trevert}{
\bsegment 
\move (1 0) \lvec (6 0) \smalldot \lvec (10 4) \move (6 0) \lvec (10 -4)
\esegment
}

\newcommand{\trekant}{
\bsegment \setunitscale{0.8}
\move (1 0) \lvec (6 0) \smalldot \lvec (16 10)
\move (6 0) \lvec (16 -10)
\move (12 6) \smalldot  \lvec (12 -6) \smalldot
\esegment
}

\newcommand{\tovert}{
\bsegment 
\move (1 0) \lvec (6 0) \smalldot \lvec (11 0)
\esegment
}

\newcommand{\tokant}{
\bsegment \setunitscale{0.8}

\move (0 -10) \lvec (0 10)
      \move (0 5) \smalldot
      \move (0 -5) \smalldot
      \move (0 0) \larc r:5 sd:-90 ed:90

\esegment
}

\newcommand{\freeEllipsis}[3]{
    \writeps {
      gsave 
      #3 rotate 
    }
    \lellip rx:#1 ry:#2
    \writeps { 
      grestore
    }
}

\newcommand{\red}{\writeps{1 0 0 setrgbcolor}}

\usepackage{amsmath}
\usepackage[all]{xy}
\SelectTips{cm}{}
\newdir{ >}{{}*!/-7pt/@{>}}
\entrymodifiers={+!!<0pt,\fontdimen22\textfont2>}

\newcommand{\drpullback}[1][dr]{\save*!/#1-1.2pc/#1:(-1,1)@^{|-}\restore}

\usepackage{mathrsfs}
\newcommand{\CC}{\mathscr{C}}

\newcommand{\upperstar}[1]{{#1}^{\raisebox{-0.25ex}[0ex][0ex]{\(\ast\)}}}
\newcommand{\lowerstar}[1]{{#1}_{\raisebox{-0.33ex}[-0.5ex][0ex]{\(\ast\)}}}
\newcommand{\lowershriek}[1]{{#1}_!}
\renewcommand{\upperstar}[1]{\Delta_{#1}}
\renewcommand{\lowerstar}[1]{\Pi_{#1}}
\renewcommand{\lowershriek}[1]{\Sigma_{#1}}

\newcommand{\fat}[1]{\mathbf{#1}}
\providecommand{\kat}[1]{\text{\textbf{\textsl{#1}}}}

\newcommand{\Set}{\kat{Set}}
\newcommand{\Grpd}{\kat{Grpd}}
\newcommand{\PrSh}{\kat{PrSh}}
\newcommand{\B}{\mathbb{B}}
\newcommand{\C}{\mathbb{C}}

\newcommand{\N}{\mathbb{N}}

\newcommand{\Hom}{\operatorname{Hom}}
\newcommand{\Aut}{\operatorname{Aut}}

\newcommand{\isleftadjointto}{\dashv}

\newcommand{\isopil}{\stackrel{\raisebox{0.1ex}[0ex][0ex]{\(\sim\)}}%
			{\raisebox{-0.15ex}[0.28ex]{\(\rightarrow\)}}}

\newcommand{\name}[1]{\ulcorner #1 \urcorner}
\newcommand{\grint}[3]{\int^{#1 \in #2} #3}

\newcommand{\h}{homotopy\xspace}

\newtheorem{taller}[thm]{$\!\!$}
\newenvironment{blanko}[1]%
{\begin{taller}{\normalfont\bfseries  #1}\normalfont}%
{\end{taller}}

\begin{document}

\thispagestyle{pageone}

\vspace*{102pt}

\begin{center}
  {\Large 
  \textbf{Data types with symmetries and \\ 
  polynomial functors over groupoids
  }\par}
  
  \vskip 22pt
  
  {\large Joachim Kock\footnote{Email:
  \href{mailto:kock@mat.uab.cat} {\texttt{\normalshape kock@mat.uab.cat}}}$^,$%
\footnote{The author has profited greatly from related collaborations with
  Andr\'e Joyal, Nicola Gambino, Anders Kock, Imma G\'alvez, Andy Tonks, 
  and notably David Gepner.
  Partial support from 
  research 
  grants 
  MTM2009-10359 
and
MTM2010-20692 
of Spain is gratefully acknowledged.}
}

\vspace{12pt}

{\small \itshape{Departament de matem\`atiques \\ 
Universitat Aut\`onoma de Barcelona \\[-6pt]
    Spain}}
    
\end{center}

{\small

   \vskip 20pt
   \hrule height 0.4pt
   \vskip 5pt

  \noindent 
  {\bf Abstract}
  \par\medbreak
  
  \noindent Polynomial functors (over $\Set$ or other locally cartesian closed
  categories) are useful in the theory of data types, where they are often
  called containers.  They are also useful in algebra, combinatorics, topology,
  and higher category theory, and in this broader perspective the polynomial
  aspect is often prominent and justifies the terminology.  For example,
  Tambara's theorem states that the category of finite polynomial functors is
  the Lawvere theory for commutative semirings \cite{Tambara:CommAlg},
  \cite{Gambino-Kock:0906.4931}.
  
  \noindent In this talk I will explain how an upgrade of the theory from sets
  to groupoids (or other locally cartesian closed $2$-categories) is useful to
  deal with data types with symmetries, and provides a common generalisation of
  and a clean unifying framework for quotient containers (in the sense of Abbott
  et al.), species and analytic functors (Joyal 1985), as well as the stuff
  types of Baez and Dolan.  The multi-variate setting also includes relations
  and spans, multispans, and stuff operators.  An attractive feature of this
  theory is that with the correct homotopical approach --- homotopy slices,
  homotopy pullbacks, homotopy colimits, etc.~--- the groupoid case looks
  exactly like the set case.
  
  \noindent After some standard examples, I will illustrate the notion of
  data-types-with-symmetries with examples from perturbative quantum field
  theory, where the symmetries of complicated tree structures of graphs play a
  crucial role, and can be handled elegantly using polynomial functors over
  groupoids.  (These examples, although beyond species, are purely combinatorial
  and can be appreciated without background in quantum field theory.)
  
  \noindent Locally cartesian closed $2$-categories provide semantics for a
  $2$-truncated version of Martin-L\"of intensional type theory.  For a
  fullfledged type theory, locally cartesian closed $\infty$-categories seem to
  be needed.  The theory of these is being developed by David Gepner and the
  author as a setting for homotopical species, and several of the results
  exposed in this talk are just truncations of $\infty$-results obtained in
  joint work with Gepner.  Details will appear elsewhere.
  
  \vskip 5pt
  

    \vskip 7pt
    \hrule height 0.4pt
    \par

  \vfill
}

\newcommand{\arxiv}[1]{\href{http://arxiv.org/pdf/#1}{ArXiv:#1}}

\section{Polynomial functors over $\Set$ and data types}
\label{intro}

\begin{blanko}{Polynomial functors in one variable.}\label{bla:1var}
  In its simplest form, a {\em polynomial functor} is an endofunctor
of $\Set$ of the form
\begin{equation}
X \mapsto \sum_{b\in B} X^{E_b} .
  \label{1var}
\end{equation}
Here the sum sign is disjoint union of sets, 
$X^{E_b}$ denotes the 
hom set $\Hom(E_b,X)$, and
$(E_b\mid b\in B)$
is a $B$-indexed family of sets, encoded 
conveniently as a single map of sets
$$
E\to B.
$$
Viewed as a data type constructor, $E\to B$ is often called a {\em container} 
\cite{Abbott,Abbott-et-al:fossacs03,Abbott-Altenkirch-Ghani:strictly-positive,Abbott-et-al:tlca03,Abbott-Altenkirch-Ghani-McBride:qotient,Altenkirch-Morris:lics09}%
; then $B$ is regarded as a set of shapes,
and the fibre $E_b$ is the set of positions in the shape corresponding to $b$.
The data to be inserted into these positions can be of any type $X$:
the polynomial functor receives a type $X$ (a set)
and returns the new more elaborate type $\sum X^{E_b}$.
Polymorphic functions
correspond to natural transformations of polynomial functors, and
these can be handled in terms of the representing sets $E \to B$ alone,
cf.~\cite{Abbott}, \cite{Gambino-Kock:0906.4931}, and \ref{NT} below.
A fundamental example is the list functor, $X\mapsto \sum_{n\in \N} X^{\fat n}$,
which to a set $X$ associates the set of lists of elements in $X$.
Here $n\in \N$ is the shape, and $\fat n$ denotes the 
$n$-element set $\{0,1,\ldots,n-1\}$ of positions in 
a length-$n$ list.
%

There is another important use of polynomial functors in type theory:
one then regards
$E\to B$ as a signature generating an algebra, namely the initial algebra for
the polynomial functor.  Initial algebras for polynomial functors are inductive
data types, corresponding to W-types 
in (extensional) Martin-L\"of type 
theory~\cite{Petersson-Synek}, \cite{Moerdijk-Palmgren:Wellfounded}.
Similarly, terminal coalgebras are coinductive data types (sometimes called 
M-types), often interpreted as programs or systems (see for 
example~\cite{Rutten:universal-coalgebra},
\cite{Hancock-Setzer:CSL2000}).
\end{blanko}

\begin{blanko}{Species and analytic functors.}
A functor is {\em finitary} when it preserves $\omega$-filtered colimits.  For a
polynomial functor this is equivalent to $E\to B$ having finite fibres. 
Let
$\B_\omega$ denote the groupoid of finite sets and bijections.  A {\em species} 
\cite{Joyal:1981} is a
functor $F:\B_\omega \to \Set$, written $S \mapsto F[S]$; the set $F[S]$ is to be
thought of as the set of $F$-structures that can be put on the set $S$.  The
{\em extension} of $F$ is the endofunctor
\begin{align}\label{anal}
  \Set \ & \longrightarrow \ \ \Set  \\
  X\  & \longmapsto \  \sum_{n\in \N} \frac{F[n] \times X^n}{\Aut(n)} \notag
\end{align}
which is the left Kan extension of $F$ along the (non-full) inclusion $\B_\omega\subset
\Set$.  A functor of this form is called {\em analytic} \cite{Joyal:foncteurs-analytiques}.
Joyal established an equivalence of
categories between species and analytic functors, and characterised analytic
functors as the finitary functors preserving cofiltered limits and weak
pullbacks ~\cite{Joyal:foncteurs-analytiques}, see also~\cite{Hasegawa}
and~\cite{Adamek-Velebil:TAC21}. 
Finitary polynomial functors are precisely the analytic functors
which preserve pullbacks strictly.  In terms of species they correspond to
those for which the symmetric group actions are free.

Monoids in species (under the operation of substitution, which corresponds to
composition of analytic functors) are precisely operads.
Many important polynomial functors
have the structure of monad.  For example, 
the list functor has a natural monad structure by concatenation of lists.
Polynomial monads equipped with a cartesian monad map to the list monad are the
same thing as non-symmetric operads~\cite{Leinster:0305049}.
More generally, finitary
polynomial monads correspond to projective operads~\cite{Kock:0807} (i.e.~such
that every epi to it splits).
\end{blanko}

\begin{blanko}{Polynomial functors in many variables.}
  Following \cite{Gambino-Kock:0906.4931}, a {\em polynomial} is a diagram of sets
\begin{equation}\label{IEBJ}
I \stackrel s \longleftarrow E \stackrel p \longrightarrow B \stackrel t 
\longrightarrow J ,
\end{equation}
and
the associated {\em polynomial functor} (or the {\em extension} of the polynomial)
is given by the composite
\begin{equation}\label{poly}
\Set_{/I} \overset{\upperstar{s}}\longrightarrow
\Set_{/E} \overset{\lowerstar{p}}\longrightarrow
\Set_{/B} \overset{\lowershriek{t}}\longrightarrow
\Set_{/J} \,,
\end{equation}
where $\upperstar{s} $ is pullback along $s$,
$\lowerstar{p} $ is the right adjoint
to pullback (called dependent product), 
and $\lowershriek{t}$ is left adjoint to 
pullback (called dependent sum).
For a map $f:B\to A$ we have the three explicit formulae
\begin{align}
\upperstar{f}  (X_a\mid a\in A) &= ( X_{f(b)}\mid b\in B)
  \label{pbk}\\[3pt]
\lowershriek{f} (Y_b \mid b\in B) &= (\underset{b\in B_a}{\textstyle{\sum}} Y_b \mid a\in A)
  \label{sum}\\
\lowerstar{f}  (Y_b \mid b\in B) &= (\underset{b\in B_a}{\textstyle{\prod}} Y_b \mid a\in A) \,,
  \label{prod}
\end{align}
giving altogether the following formula for \eqref{poly}
$$
(X_i \mid i\in I) \longmapsto (\sum_{b\in B_j} \prod_{e\in E_b} X_{s(e)}  \mid 
j\in J) ,
$$
which specialises to \eqref{1var} when $I=J=1$.

The multi-variate polynomial functors correspond to {\em indexed 
containers}~\cite{Altenkirch-Morris:lics09}, and their initial algebras
are sometimes called {\em general tree types}~\cite[Ch.~16]{Nordstrom-Petersson-Smith}.

From the abstract description in terms of adjoints, it follows that the notion 
of polynomial functor (and most of the
theory) makes sense in any locally cartesian closed category, and polynomial
functors are the most natural class of functors between slices of such 
categories.  They have been characterised intrinsically~\cite{Kock-Kock:1005.4236} 
as the local fibred right adjoints.
\end{blanko}

\begin{blanko}{Incorporating symmetries.}
  A container is a rigid data structure: it does  not allow for data
to be permuted in any way among the positions of a given shape.
In many situations it is desirable to allow for permutation, so that
certain positions within a shape become indistinguishable.
In quantum physics, the principle of indistinguishable particles
imposes such symmetry at a fundamental level.
A fundamental
example is the multiset data type,
whose extension is the functor
\begin{equation}\label{exp}
X\mapsto \sum_{n\in\N} \frac{X^{\fat n}}{\Aut(\fat n)},
\end{equation}
which is analytic but not polynomial.

In order to account for such data types with symmetries,
Abbott et al.~\cite{Abbott-Altenkirch-Ghani-McBride:qotient}  (see also 
Gylterud~\cite{Gylterud})
have extended the container formalism by adding the symmetries `by hand':
for each shape (element $b$ in $B$) there
is now associated a group of symmetries of the fibre $E_b$, and data inserted
into the corresponding positions is quotiented out by this group action. 
It is not difficult to see (cf.~also \cite{Adamek-Velebil:TAC21})
that in the finitary case, this is precisely the notion of species
and analytic functors. 

In fact it has been in the air for some time (see for example
\cite{Fiore:FOSSACS2005}, and more recently \cite{Carette-Uszkay},
\cite{Yorgey:2010}) that species should be a good framework for data type
theory.  It is the contention of the present contribution that polynomial
functors over groupoids provide a clean unifying framework: in the setting of
groupoids, the essential distinction between `analytic' and `polynomial'
evaporates (\ref{anal=poly}), and the functors can be represented by diagrams
with combinatorial content \eqref{IEBJ} just as polynomials over sets, as we
proceed to explain.

From the viewpoint of species, there are other reasons for this upgrade anyway.
In fact, it was soon realised by combinatorists that the 1985 notion of analytic
functors is not optimal for enumerative purposes: taking
cardinality simply does not yield
the exponential generating functions central to enumerative combinatorics! 
(It does so if the analytic functor is polynomial.) In
fact, the Species Book~\cite{Bergeron-Labelle-Leroux} does not mention analytic
functors at all.

The issue was sorted out by Baez and Dolan~\cite{Baez-Dolan:finset-feynman}: the
problem is that dividing out by the group action in \eqref{anal} is a bad
quotient from the viewpoint of homotopy theory, and does not behave well with
respect to cardinality.  To get the correct cardinalities, it is necessary to
use {\em homotopy quotients}, and the result is then no longer a set but a
groupoid, and the cardinality has to be {\em homotopy cardinality}.  So it is
necessary to work from the beginning with groupoids instead of sets.  Baez and
Dolan introduced {\em species in groupoids} (\ref{stuff}), dubbing them {\em
stuff types}, showed that homotopy cardinality gives the correct generating
functions, and illustrated the usefulness of the broader generality by showing
how the types needed for a combinatorial description of the quantum harmonic
oscillator are stuff types, not classical
species~\cite{Baez-Dolan:finset-feynman}.

Joint work with David Gepner closes the circle by observing that over
groupoids, species/analytic functors are the same thing as discrete finitary
polynomial functors (\ref{anal=poly});
hence the neat formalism of polynomials provides a natural unifying
framework for (quotient) containers and species.
\end{blanko}

\section{Polynomial functors over groupoids}

A groupoid is a category in which all arrows are invertible.  A useful intuition
for the present purposes is that groupoids are `sets fattened with symmetries'.
From the correct homotopical viewpoint groupoids behave very much like sets.  We
are interested in groupoids up to equivalence, and for this reason many familiar
$1$-categorical notions, such as pullback and fibre, are not appropriate, as
they are not invariant under equivalence.  The good notions are the
corresponding {\em homotopy} notions, which we briefly recall.  They can all be
deduced from the beautiful simplicial machinery developed by
Joyal~\cite{Joyal:qCat+Kan,Joyal:CRM} to generalise the theory of categories to
quasi-categories (called $\infty$-categories by Lurie~\cite{Lurie:HTT}).  Since
the $2$-category $\Grpd$ of groupoids has only invertible $2$-cells, it is an
example of a quasi-category.  From now on when we say $2$-category we shall mean
`$2$-category with only invertible $2$-cells'.

\begin{blanko}{Slices.}
  If $I$ is a groupoid, the {\em \h slice} $\Grpd_{/I}$ is the $2$-category 
  of {\em projective cones} with base $I$ (cf.~\cite{Joyal:qCat+Kan}): its  
  objects are maps $X \to I$; its arrows
are triangles with a $2$-cell
$$
\xymatrix@C=3ex{
X \ar[rr] \ar[rd] & \ar@{}[d]|{\Rightarrow} & Y \ar[ld] \\
&I&
}$$
and $2$-arrows are diagrams
$$
\xymatrix@C=3ex{
X \ar@{}[rr]|{\Uparrow}
\ar@/_0.5pc/[rr]\ar@/^0.5pc/[rr] \ar[rd] & \ar@{}[d]|{\Rightarrow^\Rightarrow}& 
Y \ar[ld] \\
&I&
}$$
commuting with the structure triangles.
More generally, if $d:T \to \Grpd$ is any diagram, there is a $2$-category
$\Grpd_{/d}$ of projective cones with base $d$.

A {\em \h terminal object} in a $2$-category $\CC$ is an object $t$ such that 
for any other object $x$, the groupoid $\CC(x,t)$ is contractible,
i.e.~equivalent to a point.
More general \h limits are defined in the usual way using \h slices:
the \h limit of a functor $d:T \to \Grpd$ is by definition a \h terminal
object in the \h slice $\Grpd_{/d}$.  
Homotopy limits are unique up to equivalence.
\end{blanko}

\begin{blanko}{Pullbacks and fibres.}
  Given a diagram of groupoids $X,Y,S$ indicated by the solid arrows,
$$
\xymatrix{
X\times_S Y \drpullback \ar@{-->}[r]\ar@{-->}[d]
&Y\ar[d]^-{g}\\X\ar[r]_-{f}&S
}
$$
the {\em \h pullback} is the \h limit, i.e.~given as a  \h terminal 
object in a a certain slice $2$-category of projective cones over the solid 
diagrams of the shape in question, and as such it is determined uniquely up to
equivalence.
A specific model is 
the groupoid $X\times_S Y$ whose objects 
are triples
$(x,y,\phi)$
 with $x \in X$, $y\in Y$ and  $\phi:fx\to gy$ an arrow of $S$, and whose
 arrows are pairs $(\alpha,\beta):(x,y,\phi)\to(x',y',\phi')$ consisting
 of
 $\alpha: x \to x'$ an arrow in $X$ and $\beta: y \to y'$ an arrow in $Y$
 such that the following diagram commutes in $S$
$$
\xymatrix{
fx\ar[r]^-\phi\ar[d]_-{f(\alpha)}&gy\ar[d]^-{g(\beta)}\\
fx'\ar[r]_-{\phi'}&gy'.
}
$$
(One should note that if $f$ or $g$ is a fibration then the na\"ive 
set-theoretic pullback is equivalent to the \h pullback.)

The \emph{\h fibre} $E_b$ of a morphism $p:E\to B$ over an object $b$ in $B$ is 
the \h pullback of $p$ along the  inclusion map
$\xymatrix{1\ar[r]^-{\name b}&B}$:
$$
\xymatrix{
E_b\drpullback\ar[r]\ar[d]
&E\ar[d]^-{p}\\1\ar[r]_-{\name b}&B.
}
$$
(Note that the homotopy fibre $E_b$ is not in general a
subgroupoid of $E$, although the map $E_b \to E$ is always faithful.
But again, if $p$ is a fibration then the set-theoretic fibre is equivalent to
the \h fibre.)
\end{blanko}

\begin{blanko}{Homotopy quotients.}
  Whenever a group $G$ acts on a set or a groupoid $X$, the
  {\em \h quotient} $X/G$ is the groupoid obtained by gluing in a path
  (i.e.~an arrow)
  between $x$ and $y$ for each $g\in G$ such that $gx=y$.
  More formally it is the total space of the Grothendieck construction 
  of the presheaf $G \to \Grpd$ that the action constitutes;
  it is a special case 
  of a \h colimit. 
  (The notation $X/\!/G$ is often used~\cite{Baez-Dolan:finset-feynman}.) 
  If $G$ acts on the point groupoid $1$, then 
  $1/G$ is the groupoid with one object and vertex group $G$.

  If $p:X\to B$ is a morphism of groupoids, for $b\in B$ the
  `inclusion' of the \h fibre $X_b \to X$ is faithful but not full in general.  But
  $\Aut(b)$ acts on $X_b$ canonically, and the \h quotient $ X_b /\Aut(b) $
  provides exactly the missing arrows, so as to make the natural map $X_b
  /\Aut(b) \to X$ fully faithful.  Since every object $x\in X$ must map to some
  connected component of $B$, we find the equivalence
\begin{align}
X &\simeq
\sum_{b\in \pi_0 B} X_b/\Aut(b) =: \grint{b}{B}{X_b},
\end{align}
expressing $X$ as the \h sum of the fibres, or equivalently as a family 
of groupoids (indexed by $\pi_0(B)$ and with a group action in each).
 Given morphisms of groupoids $Y \stackrel p \to B \stackrel f \to A$, we have
 the following `Fubini formula':
  $$
  \grint{b}{B}{Y_b} \simeq \grint{a}{A}{ \grint{b}{B_a}{Y_b}}
  $$
  \end{blanko}
which is actually the formula for the `dependent-sum' functor 
$\lowershriek{f} : \Grpd_{/B} \to \Grpd_{/A}$ given by postcomposition.
In family notation the formula reads
$$
\lowershriek{f}(Y_b \mid b\in B) \ = \ (\textstyle{\grint{b}{B_a}{Y_b}} \mid a\in A) \,,
$$
just as Formula~\eqref{sum} in the $\Set$ case.

Pullback along $f:B\to A$, denoted $\upperstar{f}$,
is right adjoint to $\lowershriek{f}$.  This means
of course \h adjoint, and amounts to a natural equivalence of mapping
groupoids
$\Grpd_{/A}(\lowershriek{f} Y, X) \simeq \Grpd_{/B}(Y,\upperstar{f} X) .$
The proof relies on the universal property of the pullback.
One may note the following formula for pullback, in family notation:
$$
\upperstar{f}(X_a\mid a\in A) \ = \ (X_{f(b)} \mid b\in B) ,
$$
again completely analogous to the $\Set$ case (Formula~\eqref{pbk}).

The $2$-category of groupoids is locally cartesian closed.
This means that the pullback functor in turn has a right adjoint
$\lowerstar{f} : \Grpd_{/B} \to \Grpd_{/A}$.  The general formula is
an end formula;
for $Y \to B$, the
fibre of $\lowerstar{f} Y$ over $a\in A$ can be described explicitly as the
mapping groupoid
$$
(\lowerstar{f} Y)_a = \Grpd_{/B}( B_a , Y) .
$$
(A more explicit formula will be derived in the discrete case below.)

\begin{blanko}{Polynomial functors.}
  A {\em polynomial} is a diagram of groupoids
  $$
I \stackrel s \longleftarrow E \stackrel p \longrightarrow B \stackrel t 
\longrightarrow J .
$$
The associated {\em polynomial functor} (or the {\em extension} of the polynomial)
is given as the composite
$$
\Grpd_{/I} \overset{\upperstar{s}}\longrightarrow
\Grpd_{/E} \overset{\lowerstar{p}}\longrightarrow
\Grpd_{/B} \overset{\lowershriek{t}}\longrightarrow
\Grpd_{/J}  .
$$
\end{blanko}

\begin{blanko}{Beck--Chevalley, distributivity, and composition.}
  Given a \h pullback square
  $$
  \xymatrix{
  \cdot \drpullback \ar[r]^{\psi} 
  \ar[d]_{\alpha} & \cdot  \ar[d]^{\beta} \\
  \cdot \ar[r]_{\varphi}& \cdot
  }
  $$
  there are natural equivalences of functors
  $$
  \lowershriek{\alpha} \circ \upperstar{\psi}  \isopil
  \upperstar{\varphi} \circ \lowershriek{\beta} 
  \qquad \text{ and } \qquad 
    \upperstar{\beta} \circ \lowerstar{\varphi}  
    \isopil \lowerstar{\psi} \circ \upperstar{\alpha} ,
    $$
   usually called the Beck--Chevalley conditions.
A more subtle feature of the theory is distributivity, which in this setting is
an equivalence saying how to distribute dependent products over dependent sums
(and which can be interpreted as a type-theoretic form of the axiom of 
choice~\cite{Martin-Loef-1984}).
We shall not need the details here.  See \cite{Gambino-Kock:0906.4931} for the 
classical case, and
Weber~\cite{Weber:1106.1983} for a deeper treatment.
The Beck--Chevalley conditions and distributivity yield a formula for
composing polynomial functors~\cite{Gambino-Kock:0906.4931}.
\end{blanko}

\begin{blanko}{Natural transformations.}\label{NT}
  Just as in the classical case~\cite{Gambino-Kock:0906.4931}, \h cartesian
  natural transformations $P'\Rightarrow P$ of polynomial functors (in one 
  variable) correspond precisely to
  \h cartesian diagrams
  $$
  \xymatrix{
  E' \ar[r]\ar[d]\drpullback & B' \ar[d]\\
  E \ar[r] & B .
  }$$
  This is an easy consequence of Beck--Chevalley.
  Showing more generally that arbitrary natural transformations are given
  essentially uniquely by diagrams
  $$
  \xymatrix{
  E' \ar[r] & B' \ar@{=}[d]\\
  \cdot \ar[u]\ar[r]\ar[d]\drpullback & B'\ar[d]\\
  E \ar[r] & B
  }$$
  is a bit more involved and depends on a \h  version of the Yoneda lemma.
  (At the time of this writing, this result is not as precise as in the
  $1$-dimensional case of \cite{Gambino-Kock:0906.4931}.)
\end{blanko}

\begin{blanko}{Spans and stuff operators.}
  Spans of groupoids are the special case of groupoid polynomials where the
  middle map is the identity (or an equivalence).  These constitute a
  categorification of matrix algebra, and were called {\em stuff operators} by
  Baez and Dolan~\cite{Baez-Dolan:finset-feynman}; they have been used to give
  groupoid models for certain aspects of Hecke algebras and Hall
  algebras~\cite{Baez-Hoffnung-Walker:0908.4305}.
\end{blanko}

\section{Exactness; combinatorial polynomial functors}

The following results from \cite{Gepner-Kock} are actually proved in the much 
richer setting
of \mbox{$\infty$-groupoids}, but the proofs work also for $1$-groupoids.  We now 
suppress
the clumsy `\h' everywhere, although of course all limits and colimits mentioned
refer to the homotopy notions.

\begin{thm} 
  (Gepner-Kock~\cite{Gepner-Kock}.)  A functor $\Grpd_{/I} \to
  \Grpd_{/J}$ is polynomial if and only if it is accessible and preserves
  conical limits.
\end{thm}
By {\em conical limit} we mean 
limit over a diagram with a terminal vertex.
Recall that a functor is accessible \cite[Ch.~5]{Lurie:HTT} when it preserves $\kappa$-filtered 
colimits for some regular cardinal $\kappa$.  The regular cardinal here is 
explicitly characterised:

\begin{prop}(\cite{Gepner-Kock})
  A polynomial functor given by $I \leftarrow E \stackrel p \to E \to J$
  preserves $\kappa$-filtered colimits if and only if $p$ has $\kappa$-compact 
  fibres.  
\end{prop}
An important case is $\kappa=\omega$.  A groupoid is $\omega$-compact
when it has finitely many components (i.e.~$\pi_0(X)$ is a finite set) and all
vertex groups are finitely presented.

\begin{blanko}{Discreteness.}
  For many data types occurring in practice (including species and all the
  examples below), although they may have symmetries, the positions in each
  shape form a {\em discrete} groupoid, i.e.~a groupoid equivalent to a set.  In the
  polynomial formalism this amounts to the middle map $p:E\to B$ having discrete
  fibres.  In this case, the dependent product formula simplifies to
$$
(\lowerstar{p} Y)_b = \prod_{e\in \pi_0(E_b)} Y_e ,
$$
  in analogy with \eqref{prod}, and hence all the formulae look exactly like the $\Set$ case.
  
  The 
  corresponding exactness condition is preservation of sifted colimits.
  A {\em $\kappa$-sifted colimit} is a colimit over a diagram $D$ whose diagonal 
  $D\to D^S$ is cofinal for every set $S$ of cardinality $< \kappa$ \cite[Ch.~5]{Lurie:HTT}.
\end{blanko}

\begin{prop}(\cite{Gepner-Kock})
  A polynomial functor given by $I \leftarrow E \stackrel p \to E \to J$
  preserves $\kappa$-sifted colimits if and only if $p$ has $\kappa$-compact 
  discrete 
  fibres.  
\end{prop}

\begin{blanko}{Combinatorial polynomial functors.}
  We call a polynomial functor $I \leftarrow E \stackrel p \to B \to J$ {\em
  combinatorial} if the fibres of $p$ are
  equivalent to finite sets (i.e.~are $\omega$-compact discrete).
\end{blanko}

\begin{blanko}{Species in groupoids (stuff types).}\label{stuff}
  A Baez-Dolan {\em stuff type} \cite{Baez-Dolan:finset-feynman}
is a map of groupoids  $F\to\B_\omega$.
We prefer the name {\em species in groupoids}.  (A classical species is
when the map has discrete fibres, or equivalently is faithful.)
Its extension is the left \h Kan extension
of $n\mapsto F_n$ along $\B_\omega \subset \Grpd$:
\begin{align*}
  \Grpd \ & \longrightarrow \  \Grpd  \\
  X \ & \longmapsto  \ \sum_{n\in \pi_0(\B_\omega)=\N} \frac{F_n \times 
  X^n}{\Aut(n)}\,.
\end{align*}
(That's a homotopy quotient of course.)

This functor is polynomial~\cite{Gepner-Kock}: the representing groupoid map is the top
row in the pullback
$$
\xymatrix{
E\ar[r]\ar[d]\drpullback & F \ar[d]\\
\B_\omega' \ar[r] & \B_\omega .
}
$$
This map has finite discrete fibres since $\B_\omega'\to\B_\omega$
has.  (Here $\B_\omega'$ is the groupoid of finite pointed sets.)
Conversely, given a groupoid polynomial $E \to F$ with finite discrete 
fibres, 
the `classifying map' $F \to \B_\omega$ (obtained since
$\B_\omega'\to \B_\omega$ classifies finite discrete fibrations)
yields a species in groupoids.
One can check that the extension of the polynomial  agrees with the 
extension of the species.
In conclusion:
\end{blanko}

\begin{prop}\label{anal=poly} (\cite{Gepner-Kock})
  Combinatorial polynomial functors $\Grpd\to\Grpd$ are the same thing as
  analytic functors (in the sense of Baez-Dolan).
\end{prop}

Combining these results we get a `Joyal theorem':

\begin{cor} (\cite{Gepner-Kock})
  A functor $\Grpd\to\Grpd$ is analytic (in the sense of Baez-Dolan)
  if and only if it preserves $\omega$-sifted 
  colimits and conical limits.
\end{cor}

\begin{blanko}{Generalised species.} 
  The relationship between polynomial functors and the generalised species of
  \cite{Fiore-Gambino-Hyland-Winskel:Esp} has been sketched by Gambino and the
  author (unpublished).  A generalised species depends on two categories $I$ and
  $J$, and has as extension a generalised analytic functor $\PrSh(I) \to
  \PrSh(J)$; this generalises the 1985 notion but not the Baez-Dolan notion.  If
  $I$ and $J$ are groupoids, these generalised analytic functors correspond to
  the {\em `classical'} extension of combinatorial polynomials over groupoids,
  i.e.~involving $\pi_0$ on quotients. 
\end{blanko}

\begin{blanko}{Examples.}
  Groupoid polynomials encode data types in groupoids.  For example,
  $\B'_\omega \to \B_\omega$ encodes the multiset data type: the groupoid $\B_\omega$
  of finite sets and bijections is the
  groupoid of shapes --- the shape of a multiset is really the set indexing its 
  elements, not just its size.  There are $\N$-many isoclasses; 
  the isomorphisms should be interpreted as propositional equality.
  The fibre over $S\in \B_\omega$ is the discrete groupoid of
  positions in $S$, i.e.~a uniform prescription of positions in multisets 
  indexed by $S$. Indeed,  since $\B'_\omega \to \B_\omega$ is a fibration,
  the fibre is canonically identified with the set $S$ itself --- note its natural
  $\Aut(S)$-action.  The discreteness of the fibre means that propositional
  equality
  between positions can be regarded as definitional equality.
The extension of this quotient container is naturally an endofunctor
$\Grpd\to\Grpd$.  But one obtains an endofunctor $\Set\to\Set$ (in this case 
precisely \eqref{exp}) by
precomposing with the natural inclusion $\Set\to\Grpd$ and postcomposing
with $\pi_0:\Grpd\to\Set$.  The first is harmless.  The second corresponds
to collapsing all isomorphisms to identity, i.e.~interpreting propositional
equality as definitional equality.  If the argument is a set, the only 
collapse is the passage from homotopy quotient to na\"ive quotient (of
actions on sets).

The data type of cyclic lists is groupoid polynomial, represented by 
$\C'_\omega\to\C_\omega$,
where $\C_\omega$ is the groupoid of finite cyclically ordered sets, and $\C'_\omega$ is the
groupoid of pointed cyclically ordered finite sets.  
From \ref{bla:1var}, the list data type is represented by $\N'\to\N$, interpreted as
the groupoids of linearly ordered finite 
sets and pointed ditto.  The diagram of groupoids
\[
\xymatrix{
\N' \ar[r]\ar[d]\drpullback & \N \ar[d] \\
\C'_\omega \ar[r]\ar[d]\drpullback & \C_\omega \ar[d] \\
\B_\omega' \ar[r] & \B_\omega 
}\]
now represents
the cartesian natural transformations, or polymorphic functions,
from lists to cyclic lists to multisets.
\end{blanko}

\section{Trees}

W-types in Martin-L\"of type theory correspond to initial algebras of polynomial 
functors (cf.~\cite{Moerdijk-Palmgren:Wellfounded} and \cite{Gambino-Hyland}
for the extensional case, and
\cite{Awodey-Gambino-Sojakova:1201.3898}
for the fully intensional case).  The initial algebra for $1+P$ can also be 
described as the set of operations for the free monad on $P$, which in turn
is the set of {\em $P$-trees}.
$P$-trees (for $P$ a polynomial functor over 
$\Set$ or any lccc) are always rigid, i.e.~have no symmetries.
Abstract trees, on the other hand, admit symmetries, so they are
not $P$-trees for any $\Set$-polynomial functor $P$, and they are neither W-types nor
containers in the classical sense.
Instead, according to \cite{Kock:0807}, abstract trees {\em are}
themselves polynomial functors.
It is convenient to take the following
characterisation of trees as a definition:

\begin{blanko}{Trees.}\label{polytree-def}
  (\cite{Kock:0807}) A
  {\em (finite) tree} is a diagram of finite sets
    $$
    A \stackrel s \longleftarrow  M \stackrel p \longrightarrow  N  \stackrel 
    t \longrightarrow  A
$$
satisfying the following three conditions:
  
  (1) $t$ is injective
  
  (2) $s$ is injective with singleton complement (called  {\em 
  root} and denoted $1$).
  
  \noindent With $A=1+M$, 
  define the walk-to-the-root function
  $\sigma: A \to A$ by $1\mapsto 1$ and $e\mapsto t(p(e))$ for
  $e\in M$. 
  
  (3)  $\forall x\in A : \exists k\in \N : \sigma^{k}(x)=1$.
  
  The elements of $A$ are called {\em edges}.  The elements of $N$
  are called {\em nodes}.  For $b\in N$, the edge $t(b)$ is called
  the {\em output edge} of the node.  That $t$ is injective is just to
  say that each edge is the output edge of at most one node.  For
  $b\in N$, the elements of the fibre $M_b$ are
  called {\em input edges} of $b$.  Hence the whole set
  $M=\sum_{b\in N} M_b$ can be thought of as the set of
  nodes-with-a-marked-input-edge, i.e.~pairs $(b,e)$ where $b$ is a
  node and $e$ is an input edge of $b$.  The map $s$ returns the
  marked edge.  Condition (2) says that every edge is the input edge
  of a unique node, except the root edge.
  Condition (3) says that if you walk towards the root, in a finite 
  number of steps you arrive there.
  The edges not in the image of $t$ are called {\em leaves}.
\end{blanko}

\begin{blanko}{Decorated trees: $P$-trees} (\cite{Kock:0807}; see also 
  \cite{Kock:1109.5785,Kock:graphs-and-trees,Kock-Joyal-Batanin-Mascari:0706})
  An efficient way of encoding and manipulating decorations of trees
  is in terms of polynomial endofunctors.  Let $P$ be a polynomial 
  endofunctor given by $I \overset d\leftarrow E \overset q\to B \overset c\to I$.
  A {\em $P$-tree} is a diagram
  \begin{equation}\label{Ptree}
  \xymatrix @!C=16pt {
  A\ar[d] 
  & M\ar[l]  
  \drpullback\ar[r] \ar[d]& N \ar[d] \ar[r] 
  &A\ar[d] \\
  I & \ar[l] E  \ar[r] & B \ar[r]&I \,, \\
}\end{equation}
  where the top row is a tree.  The squares are commutative up to
  isomorphism, and it is important that the $2$-cells be specified as part of
  the structure.  Unfolding the definition, we see that a $P$-tree is a tree
  whose edges are decorated in $I$, whose nodes are decorated in $B$, and with
  the additional structure of an equivalence $M_n \simeq E_b$ for each node $n \in
  N$ with decoration $b \in B$ (this is essentially just a bijection, since the 
  fibres are discrete), an iso in $I$ between the decoration of an edge
  $m\in M_n$ and the corresponding $d(e)$, and finally an iso in $I$ between the
  decoration of the output edge of $n$ and $c(b)$.
\end{blanko}

\begin{blanko}{Examples of $P$-trees.}
  Natural numbers are $P$-trees for the identity monad $P(X)=X$, and
  are also the set of operations of the list monad.
  Planar finite trees are $P$-trees for $P$
  the list monad, and are also the set of operations of the free-non-symmetric-operad 
  monad~\cite{Leinster:0305049}.  These two examples are the first entries of a canonical sequence
  of inductive data types underlying several approaches to higher
  category theory, the {\em opetopes}: opetopes in dimension $n$
  are $P$-trees for
  $P$ a $\Set$-polynomial functor whose operations are 
  $(n-1)$-opetopes~\cite{Kock-Joyal-Batanin-Mascari:0706};  hence opetopes are 
  higher-dimensional trees.

  Abstract finite trees are $P$-trees for
  the multiset functor
  $
  1 \leftarrow \B'_\omega \to \B_\omega \to 1,
  $
  but cannot be realised as $P$-trees for any $\Set$-polynomial $P$.
\end{blanko}

\begin{blanko}{Trees of Feynman graphs.}
  In the so-called BPHZ renormalisation of perturbative quantum field theories,
  one is concerned with nestings of \linebreak 1-particle irreducible (1PI)
  Feynman graphs, i.e.~graphs~\cite{Joyal-Kock:0908.2675} for which no single
  edge removal disconnects.  Kreimer~\cite{Kreimer:9707029} discovered that the
  BPHZ procedure is encoded in a Hopf algebra of (non-planar) rooted trees,
  expressing the nesting of graphs.
  
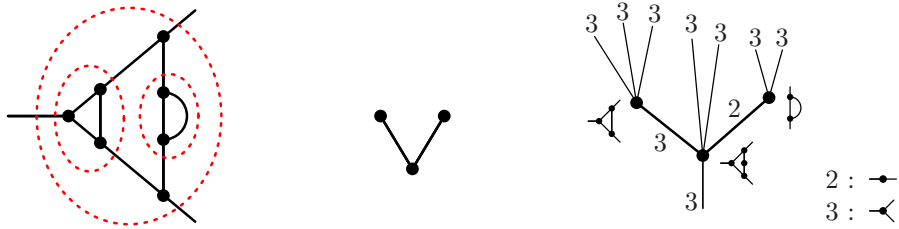
\begin{center}
  \begin{texdraw}
  \bsegment
	\linewd 1
      \move (-3 0) \lvec (20 0) \Onedot \lvec (68 40)
      \move (56 30) \Onedot \lvec (56 -30) \Onedot
      \move (32 10) \Onedot \lvec (32 -10) \Onedot
      \move (20 0) \lvec (68 -40)

      \move (56 9) \Onedot
      \move (56 -9) \Onedot
      \move (56 0) \larc r:9 sd:-90 ed:90

      \red
      \lpatt (1 3)
      \move (28 -1) \freeEllipsis{13}{20}{0}
      \move (58 0) \freeEllipsis{11}{16}{0}
      \move (43 0) \freeEllipsis{35}{41}{0}

  \esegment

  \move(150 -20)
  \bsegment
  	\linewd 1
  \move (0 0) \Onedot \lvec (-12 20) \Onedot
  \move (0 0) \lvec (12 20) \Onedot
  \esegment

\move (260 -35)

  \bsegment
  
    \move (0 0) 
    \lvec (0 20) \Onedot \lvec (-4 64) 
    \move (0 20) \lvec (6 62)
    \move (0 20) \linewd 1 
    \lvec (25 42) \linewd 0.5
    \Onedot \lvec (20 60) \move (25 42) \lvec (30 60)

    \move (0 20) \linewd 1 \lvec (-25 40) \linewd 0.5
 
    \bsegment 
      \move (0 0) \Onedot \lvec (-16 25) \move (0 0) \lvec (-5 30)
      \move (0 0) \lvec (6 29)
    \esegment

\move (6 17) \trekant 
\setunitscale{0.8} \rmove (12 0) \smalldot \setunitscale{1}
\move (-44 33) \trekant
\move (33 38) \tokant

\htext (-4 3){{\footnotesize $3$}}
\htext (-16 25){{\footnotesize $3$}}
\htext (-4 70){{\footnotesize $3$}}
\htext (7 67){{\footnotesize $3$}}

\htext (-42 70){{\footnotesize $3$}}
\htext (-30 75){{\footnotesize $3$}}
\htext (-18 74){{\footnotesize $3$}}

\htext (20 65){{\footnotesize $3$}}
\htext (30 65){{\footnotesize $3$}}
\htext (12 38){{\footnotesize $2$}}

\move (52 11)
\bsegment
\htext (0 0){{\footnotesize $2$ :}} \move (10 0) \tovert
\htext (0 -12){{\footnotesize $3$ :}} \move (10 -12) \trevert
\esegment
\esegment

\end{texdraw}
\end{center}
In the picture the combinatorial tree in the middle expresses the nesting of
1PI subgraphs on the left; such trees are sufficient in Kreimer's Hopf-algebra
approach to BPHZ, but do not capture the symmetries of the graph.
To this end, further
decoration is needed in the tree, as partially indicated on the right.
First of all, each node in the tree should be decorated by the 1PI graph
it corresponds to in the nesting, and second,
the tree should have leaves (input
slots) corresponding to the vertices of the graph. 
The decorated tree should be 
regarded as a
recipe for reconstructing the graph by inserting the decorating graphs into
the vertices of the graphs of parent nodes.  The numbers on the edges
indicate the  type constraint of each substitution: the outer interface of
a graph must match the local interface of the vertex it is substituted into.
But the type constraints on the tree decoration are not enough to reconstruct the
graph, because for example the small graph 
\raisebox{-5pt}{\begin{texdraw}\trekant\end{texdraw}} decorating
the left-hand node could be substituted into various different vertices of the 
graph
\raisebox{-5pt}{\begin{texdraw}
  \trekant  \setunitscale{0.8} \rmove (12 0) \smalldot \setunitscale{1}
\end{texdraw}}.

  The solution found in \cite{Kock:graphs-and-trees} is to consider $P$-trees,
  for $P$ the polynomial endofunctor given by $I \stackrel s\leftarrow E
  \stackrel p\to B \stackrel t\to I$, where $I$ is the groupoid of interaction
  labels for the theory (in this case the one-vertex graphs
  \raisebox{2pt}{\begin{texdraw} \tovert \end{texdraw}} and
  \raisebox{-1pt}{\begin{texdraw} \trevert \end{texdraw}}\ ) and $B$ is the
  groupoid of connected 1PI graphs of the theory, and $E$ is the groupoid of such 1PI
  graphs with a marked vertex.  The map $s$ returns the one-vertex subgraph at
  the mark, $p$ forgets the mark, and $t$ returns the outer interface of the
  graph, i.e.~the graph obtained by contracting everything to a point, but
  keeping the external lines.  A $P$-tree is hence a diagram like \eqref{Ptree}
  with specified $2$-cells.  These $2$-cells carry much of the structure: for
  example the $2$-cell on the right says that the 1PI graph decorating a given
  node must have the same outline as the decoration of the outgoing edge of the
  node --- or more precisely, and more realistically: an isomorphism is specified (it's
  a bijection between external lines of one-vertex graphs).  Similarly, the
  left-hand $2$-cell specifies for each node-with-a-marked-incoming-edge $x'\in
  M$, an isomorphism between the one-vertex graph decorating that edge and the
  marked vertex of the graph decorating the marked node $x'$.  Hence the
  structure of a $P$-tree is a complete recipe not only for which graphs should
  be substituted into which vertices, but also {\em how}: specific bijections
  prescribe which external lines should be identified with which lines in the receiving
  graph.
  In fact, {\em there is an equivalence of groupoids between nested graphs and
  $P$-trees}~\cite{Kock:graphs-and-trees}.  This is exploited
  in~\cite{Galvez-Kock-Tonks:FdB}
  to establish algebraic identities concerning 
  graphs by interpreting them as
  \h cardinalities of equivalences of groupoids of decorated trees.
  
  Notice that the polynomial functor $P$ is combinatorial, since each
  graph has a discrete finite set of vertices.  It is not a species in the 
  classical sense though: the classifying map $B \to \B_\omega$ sends a graph
  to its set of vertices, and since a graph may have nontrivial automorphisms that fix 
  every
  vertex, this map does
  not have discrete fibres.
\end{blanko}

\section{Outlook}

  A $2$-category is called {\em locally cartesian closed} when for every arrow
  $f:B\to A$, we have the string of adjoint functors
  $
  \lowershriek{f} \isleftadjointto \upperstar{f} \isleftadjointto \lowerstar{f} .
  $  This structure formally implies the Beck-Chevalley equivalences and 
  distributivity, which are the minimal requirements for a reasonable theory of 
  polynomial functors.  The theory of strength can be copied almost verbatim from 
  \cite{Gambino-Kock:0906.4931},
  and it seems that the
  representation theorem of \cite{Gambino-Kock:0906.4931} also carries over.

While locally cartesian closed categories provides semantics for an extensional
version of Martin-L\"of type theory~\cite{Seely:lccc},
\cite{Clairambault-Dybjer:1112.3456}, and locally cartesian closed $2$-categories capture
some $2$-truncated version (\cite{Hofmann-Streicher:98},
\cite{Garner:0808.2122}), recent insight of Homotopy Type Theory
strongly
suggests that in the long run, the case of $\infty$-groupoids and other locally
cartesian closed $\infty$-categories will be the real meat for type theory.
Large parts of the $\infty$-theory of polynomial functors,
as well as aspects of the theory of 
locally cartesian closed 
$\infty$-categories geared towards Homotopy Type Theory
have already been
worked out in joint work with David Gepner, and 
will appear
elsewhere~\cite{Gepner-Kock},~\cite{Gepner-Kock:univalence}.
Nevertheless the groupoid case is interesting in its own right,
since
it already covers important applications: in particular
for many purposes of combinatorial nature, $1$-groupoids are all that is
needed in order to handle symmetry issues.  Time will tell whether for
purposes of program semantics the groupoid level is enough too
--- otherwise it is a good stepping stone into the $\infty$-world.

\end{document}